# DISCOVERING POTENTIAL USER BROWSING BEHAVIORS USING CUSTOM-BUILT APRIORI ALGORITHM

Sandeep Singh Rawat[1] and Lakshmi Rajamani[2]

[1] Department of Computer Science & Engineering,
Guru Nanak Institute of Technology, Ibrahimpatnam,
Andhra Pradesh 501506, India
sandeep4578@gmail.com

[2] Department of Computer Science & Engineering,
College of Engineering, Osmania University,
Hyderabad, Andhra Pradesh, India
drlakshmiraja@gmail.com

## ABSTRACT

*Most of the organizations put information on the web because they want it to be seen by the world. Their goal is to have visitors come to the site, feel comfortable and stay a while and try to know completely about the running organization. As educational system increasingly requires data mining, the opportunity arises to mine the resulting large amounts of student information for hidden useful information (patterns like rule, clustering, and classification, etc). The education domain offers ground for many interesting and challenging data mining applications like astronomy, chemistry, engineering, climate studies, geology, oceanography, ecology, physics, biology, health sciences and computer science.*

*Collecting the interesting patterns using the required interestingness measures, which help us in discovering the sophisticated patterns that are ultimately used for developing the site. We study the application of data mining to educational log data collected from Guru Nanak Institute of Technology, Ibrahimpatnam, India. We have proposed a custom-built apriori algorithm to find the effective pattern analysis. Finally, analyzing web logs for usage and access trends can not only provide important information to web site developers and administrators, but also help in creating adaptive web sites.*

## KEYWORDS

*Association Rule Mining, Data Mining, and Web log*

## 1. INTRODUCTION

The enormous content of information on the World Wide Web makes it an obvious candidate for data mining research. Application of data mining techniques to the World Wide Web is referred to as Web Mining. This area of Web Mining can be divided into three sub areas - Web Content Mining, Web Structure Mining and Web Usage Mining. Web mining is a technology on finding out implicit pattern, $p$, from vast Web document structures and muster, $C$. If we see $C$ as export and $p$ as import, then, the process of Web mining can be thought of as a mapping from import to export: $\xi : C \rightarrow p$. Accurate Web usage information could help to attract new customers, retain current customers, and improve cross marketing/sales, effectiveness of promotional campaigns, tracking leaving customers and find the most effective logical structure for their Web space. User profiles could be built by combining users' navigation paths with other data features, such as page viewing time, hyperlink structure, and page content. At present, there are about three kinds of search engines in Internet. The one is search engine based on artificial, such as Yahoo. The other is search engine based on Robot, for example, AltaVista, Lycos and Excite. And the third is meta-search engine, for instance, Byte-search, MetaCrawler and Ixquick, etc. Although the present search engines have taken great conveniences for people's searching information, it has little good effect. On the other hand, data mining technology is becoming more and more mature after many years' development. And it may use computer forwardly to distill valuable data modes from enormous data to meet people' different demands. So currently it is an important task to import data mining into Web information retrieval. Web usage mining is a process of picking up information from user how to use web sites. Web content mining is a process of picking up information from texts, images and other contents. Web structure mining is a process of picking up information from linkages of web pages, such as Table 1.

Table 1. The relationship among the different areas of Web Mining

| Type | Structure | Form | Object | Collection |
|------|-----------|------|--------|------------|
| Usage | Accessing | Click | Behaviour | Logs |
| Content | Pages | Text | Index | Pages |
| Structure | Map | Hyperlinks | Map | Hyperlinks |

The access data of the users visiting a given web site is provided by the Server Log Files. They provide details about file requests to a web server and the server response to those requests. In the access log, which is the main log file, each line describes the source of a request, the file requested, the date and time of the request, the content type and length of the transferred file, and other data such as errors and the identity of referring pages.

The rest of the paper is organized as follows. Section II discusses the related work. Section III explains the solution framework used for generation of rules for educational web site analysis. Section IV gives some intuition about the observed results and explains about the how the different co-relations and rules are generated and presents the simulations that demonstrate comparisons between the apriori algorithm and the custom-built apriori. Finally, section V concludes and future enhancements of the system are mentioned.

## 2. RELATED WORK

The wide usage of the Internet in various fields has increased the automatic extraction of the log data from the web sites. The usage of data mining techniques on the data collected from the web helps us pattern selection, which acts as a traditional way of decision-making tools. Web usage mining is the application of the data mining techniques on the web-collected data, which is already present in the form of various patterns. Web usage mining is presented on secondary data such as (user name, ip address, date and time, their type of browsers used, type of URL used to view the site etc.) which is deduced from the interactions of the users in between the web sessions. Wang tong HE Pi-lian in their paper, "*Web Log Mining by an Improved AprioriAll algorithm*" showed that the possibility and importance about applying Data Mining in Web log mining and showed some problems in the conventional searching engines. Then it offers an improved algorithm based on the original AprioriAll algorithm, which has been used in Web, logs mining widely. Test results show the improved algorithm has a lower complexity of time and space [1]. Vic Ciesielski and Anand Lalani in their paper *"Data Mining of Web Access Logs from an Academic Web Site"* used a general-purpose data-mining tool to determine whether we can find any 'golden nuggets' in the web access logs of a large academic web site. They discovered several nuggets, the most significant being that a major difference between visitors from within Australia and visitors from outside Australia generally arrive via search engines and are interested in information about postgraduate courses [2]. Gui-Rong Xue et al. proposed a paper on "*Log Mining to Improve the Performance of Site Search*". This paper [3] proposes a novel re-ranking method based on user logs within websites. With the help of website taxonomy, they mine for generalized association rules and abstract access patterns of different levels. Mining results are subsequently used to re-rank the retrieved pages.

Maristella Agosti and Giorgio Maria Di Nunzio published a paper on *"Web Log Mining: A Study of User Sessions"*. This study [4] reports on initial findings on a specific aspect that is highly relevant for personalization services like study of web user sessions. Mike Thelwall in his paper [5] *"Web Log File Analysis: Backlinks and Queries"* stated that web log files are a useful source of information about visitor site use, navigation behavior and to some extent demographics. Wen-Chen Hu et al. had their research over web usage mining and published a paper [6] on *"World Wide Web Usage Mining Systems and Technologies"*. According to this web usage mining is used to discover interesting user navigation patterns and can be applied to many real-world problems, such as improving Web sites/pages, making additional topic or product recommendations, user/customer behavior studies, etc. Zhenglu Yang et al. published a paper on *"An Effective System for Mining Web Log"*. In this paper [7], they proposed an effective web log-mining system consists of data preprocessing, sequential pattern mining and visualization. Osmar R. Zaiane proposed a paper on *"Web Usage mining for a Better Web-Based Learning Environment"*. In this paper [8], they discussed some data mining and machine learning techniques that could be used to enhance web-based learning environments for the educator to better evaluate the leaning process. Magdalini Eirinaki and Michalis Vazirgiannis published a paper titled *"Web Mining for Web Personalization"*. In this paper [9] they presented a survey of the use of web mining for web personalization.

Yongjian Fu and Ming-Yi Shih published a paper titled *"A Framework for Personal Web Usage Mining"*. In this paper [10], they proposed a framework to mine Web usage data on client side, or personal Web usage mining, as a complement to the server side Web usage mining. Renata Ivancsy and Istvan Vajk proposed a paper on *"Different Aspects of Web Log Mining"*. In this paper [11] three of the most important approaches are introduced for web log mining. All the three methods are based on the frequent pattern mining approach. Federico Michele Facca and Pier Luca Lanzi published a paper on *"Recent Developments in Web Usage Mining Research"*. In their terms Web Usage Mining is the area of Web Mining, which deals with the extraction of interesting knowledge from logging information produced by web servers [12]. Jaideep Srivastava et al.published a paper titled *"Web Usage Mining: Discovery and Applications of Usage Patterns from Web Data"*. According to this paper web usage mining is the application of data mining techniques to discover usage patterns from Web data, in order to understand and better serve the needs of Web-based applications [13]. Alzennyr da Silva et al. published a paper *"Mining Web Usage Data for Discovering Navigation Clusters"*. According to them, the analysis of a web site based on usage data is an important task as it provides insight into the organization of the site and its satisfactory results regarding user needs [14]. Ramakrishnan Srikant and Yinghui Yang published their paper on *"Mining Web Logs to Improve Website Organization"*. This paper states that many websites have a hierarchical organization of content. In this paper [15], they proposed an algorithm to automatically find pages in a website whose location is different from where visitors expect to find them. Sujni Paul [16] in her paper proposed an Optimization Distributed Association Rule mining algorithm for geographically distributed data is used in parallel and distributed environment so that it reduces communication costs. The response time is calculated in this environment using XML data. Keshavamurthy B.N. , Mitesh Sharma and Durga Toshniwal [17] in their paper "Efficient Support Coupled Frequent Pattern Mining Over Progressive Databases", proposed novel approach efficiently mines frequent sequential pattern coupled with support using progressive mining tree.

## 3. CUSTOM-BUILT ALGORITHM

A server log is a log file (or several files) automatically created and maintained by a server of activity performed by it. A typical example is a web server log which maintains a history of page requests. These files are usually not accessible to general Internet users, only to the webmaster or other administrative person. A statistical analysis of the server log may be used to examine traffic patterns by time of day, day of week, referrer, or user agent. Efficient web site administration, adequate hosting resources and the fine tuning of sales efforts can be aided by analysis of the web server logs. In the generated analysis we try to prepare an analysis that can be successfully be utilized by the web site developer in order to improve his/her web site effectively. we tried to avoid all the unnecessary details such as browser details and included all the required or needed details such as analysis about the visitors, complete hits, daily analysis, searched files etc. We have generated the frequent item sets from the given database, which solves both the specified problems. First problem is to find those item sets whose occurrences exceed a predefined threshold in the database; those item sets are called frequent or large item sets. Here we are capable generating the frequent item sets by specifying the threshold as maximum number of hits. The second problem is to generate association rules from those large item sets with the constraints of minimal confidence. Suppose one of the large item sets is $L_k$, $L_k$ = {$I_1$, $I_2$, … , $I_k$}, association rules with this item sets are generated in the following way: the first rule is {{$I_1$, $I_2$, … , $I_{k-1}$}…{$Ik$}, by checking the confidence this rule can be determined as interesting or not. This problem is also solved as we specified the % of total (support) through which we can determine if the item set generated is interesting or not. See the figure 1 for the pseudo code for custom-built apriori algorithm.

```
Procedure apriori(ipadd,url)

for each item set of distipadd y

{ count=0;

for each item set of ipadd x

{ if(distip[y].equals(ipadd[x])){

if successful {

count++;

counta[k]=count;

ipu[k]=url[x];   // join step

//out.println(ipu2[k3]+y+"f");

k++; } }

if not successful

remove the itemset  //prune step

} }
```

Figure 1. Pseudo code for custom-built apriori algorithm

## 4. RESULT AND RULE GENERATION

We categorized this information in such a way that the user feels easier to understand the study made and make the necessary reforms in the site as required. Different categorizations made in the analysis are:

- General statistics
- Access statistics
- Co-relations

## 4.1. General Statistics

General statistics consists of all the general or summarized analysis which gives us the complete overview of the different fields present in the log file. This generalized study gives us the perfect count of the different fields such as

- Total number of hits
- Total number of visitors
- Different errors
- Successful visits
- Incomplete visits
- Error reports etc

This analysis completely is based on the status codes field; status codes of different numbers have different importance of their own through which it makes the job easier for the developer or the user to understand the analysis, see the figure 2. We also try to acquire the related information as to how are they reaching the specific web sited so, as by analyzing the entire available patterns in the log file these patterns can be further used in the pattern discovery in order to produce the results.

**GENERAL STATISTICS**

| | |
|---|---|
| TOTALNO OF HITS | 4776 |
| TOTALNO OF SUCCESSFUL HITS | 2717 |
| TOTALNO OF INCOMPLETE HITS | 2059 |
| PAGE VIEWS | 1029 |
| IMAGE VIEWS | 1298 |
| FILE DOWNLOADS | 59 |

**PER DAY ANALYSIS**

| DAY | HITS | SUCCESSFUL | INCOMPLETE |
|---|---|---|---|
| 27/Nov/2008 | 2504 | 1315 | 1189 |
| 28/Nov/2008 | 2272 | 1402 | 870 |
| AVERAGE HITS PER DAY | 2388 | | |

**POPULAR BROWSERS**

| BROWSER | | COUNT |
|---|---|---|
| Mozilla/4.0 | 2904 | |
| Mozilla/5.0 | 1791 | |

**ERROR REPORTS FOR PAGE ACCESS**

| | |
|---|---|
| REQUEST NOT FOUND | 1516 |

Figure 2. Output generated by general statistics

## 4.2. Access Statistics

Similar to general statistics in this access statistics, see figure 3, we try to specify the more detailed analysis, which gives us the access statistics of different users depending upon their different allocated ip address and url users. This access statistics consist of the details of both the successful and unsuccessful hits based on:

- ip address
- url

**POPULAR VISITS**

| IPADDRESS | HITS | INCOMPLETE HITS | % OF TOTAL |
|---|---|---|---|
| 117.195.200.161 | 80 | 28 | 0.65 |
| 117.195.228.4 | 31 | 4 | 0.87096775 |
| 117.199.240.105 | 57 | 1 | 0.98245615 |
| 119.235.49.2 | 492 | 243 | 0.50609756 |
| 121.246.168.221 | 102 | 18 | 0.8235294 |
| 121.246.168.223 | 95 | 65 | 0.31578946 |
| 121.246.89.189 | 64 | 21 | 0.671875 |
| 121.247.224.19 | 52 | 23 | 0.5576923 |
| 122.169.139.112 | 54 | 18 | 0.6666667 |
| 122.169.189.62 | 87 | 24 | 0.7241379 |
| 122.169.218.27 | 146 | 94 | 0.3561644 |
| 122.252.229.174 | 49 | 23 | 0.53061223 |
| 125.17.79.67 | 144 | 39 | 0.7291667 |
| 202.133.60.6 | 36 | 8 | 0.7777778 |
| 220.225.18.213 | 92 | 51 | 0.4456522 |
| 59.162.210.149 | 78 | 14 | 0.82051223 |
| 59.93.112.198 | 55 | 11 | 0.8 |
| 59.93.49.180 | 32 | 6 | 0.8125 |
| 59.93.49.40 | 94 | 63 | 0.32978722 |
| 59.93.78.104 | 336 | 224 | 0.33333334 |
| 59.93.80.148 | 92 | 20 | 0.7826087 |

Figure 3. Output generated for urls in access statistics

## 4.3. Co-relations

We try to generate the different rules among the specified item sets. The different item sets present in our analysis are:

- ip address
- url
- path

The different rules that are formed with these combinations are:

- ipadd→url
- url→path
- Ipadd→path
- Ipadd→url→path

The ipadd→url relation for example; collect all the distinct ip addresses which have visited or completed their visits successfully and then we collected the successful urls that have completed their paths successfully using the specified ip address, see the figure 4.

**POPULAR URL**

| IPADDRESS | URL | HITS | INCOMPLETE | % OF TOTAL |
|---|---|---|---|---|
| 117.195.200.161 | /combined.pdf | 3 | 0 | 1.0 |
| 117.195.228.4 | /combined.pdf | 3 | 0 | 1.0 |
| 117.198.149.101 | /combined.pdf | 3 | 0 | 1.0 |
| 117.199.240.105 | /combined.pdf | 3 | 0 | 1.0 |
| 118.94.230.71 | /combined.pdf | 3 | 0 | 1.0 |
| 118.94.243.55 | /combined.pdf | 3 | 0 | 1.0 |
| 118.95.122.188 | /combined.pdf | 3 | 0 | 1.0 |
| 119.235.49.2 | / | 11 | 5 | 0.54545456 |
| 119.235.49.2 | /atten.html | 3 | 0 | 1.0 |
| 119.235.49.2 | /atten_files/arrow.gif | 3 | 0 | 1.0 |
| 119.235.49.2 | /atten_files/menu.js | 3 | 0 | 1.0 |
| 119.235.49.2 | /atten_files/page.css | 3 | 0 | 1.0 |
| 119.235.49.2 | /atten_files/text.css | 3 | 0 | 1.0 |
| 119.235.49.2 | /atten_files/title.jpg | 3 | 0 | 1.0 |
| 119.235.49.2 | /atten_files/top/br.css | 3 | 0 | 1.0 |
| 119.235.49.2 | /atten_files/top/br.js | 3 | 0 | 1.0 |
| 119.235.49.2 | /atten_files/vignantz.ac.htm | 4 | 1 | 0.75 |
| 119.235.49.2 | /combined.pdf | 4 | 0 | 1.0 |
| 119.235.49.2 | /index_files/arrow.gif | 9 | 4 | 0.5555556 |
| 119.235.49.2 | /index_files/best.jpg | 12 | 7 | 0.41666666 |
| 119.235.49.2 | /index_files/menu.js | 12 | 5 | 0.5833333 |

Figure 4. Output generated for the generated co-relation

### 4.4. Comparisons

We try to compare the existing apriori algorithm with the custom-built apriori algorithm. See the table 2 for step by step comparisons.

Table 2. Comparisons between apriori and proposed algorithm.

| APRIORI Algorithim | Custom built APRIORI Algorithm |
|---|---|
| 1) Procedure apriori-gen(Lk-1:freq(k-1) itemset<br><br>for each itemset l1 belongs Lk-1<br><br>For each itemset l2 belongs lk-1<br><br>if(l1[1]=l2[1]^(l1[2]=l2[2]^…) then<br><br>{<br><br>c=l1*l2;  **//join step**<br><br>if has-infreq-subset(c,Lk-1)<br><br>delete c  **// prune step**<br><br>else add c to Ck;<br><br>}<br><br>Return Ck | 1) Procedure apriori(ipadd,url)<br><br>for each item set of distipadd y<br><br>{ count2=0;<br><br>for each item set of ipadd x<br><br>{ if(distip[y].equals(ipadd[x])){<br><br>if successful {<br><br>count2++;<br><br>counta[k4]=count2;<br><br>ipu2[k5]=url[x];  **// join step**<br><br>//out.println(ipu2[k3]+y+"f");<br><br>k5++; } }<br><br>if not successful<br><br>remove the itemset  **//prune step**<br><br>} |
| 2) Scan D for count of each candidate and compare support count with min supp | 2) Frequent itemsets generated in the general statistics |

| Item | Support |
|---|---|
| 1 | 3 |
| 2 | 6 |
| 3 | 4 |
| 4 | 5 |

| Item | Hits |
|---|---|
| *ipadd* | 35 |
| *url* | 48 |
| *path* | 24 |

<table>
<tr><td colspan="2">

3) Generate C2 candidates from L1. Scan D for count of each candidate. Compare

| Item | Support |
|------|---------|
| {1,2} | 3 |
| {1,3} | 2 |
| {1,4} | 3 |

candidate supp

</td><td colspan="2">

3) Generate co-relations from the above statistics and scan for the count and compare the threshold count of hits

| Item | Hits |
|------|------|
| *ipadd->url* | 25 |
| *url->path* | 38 |
| *ipadd->path* | 27 |

</td></tr>
</table>

| 3) Generate C2 candidates from L1. Scan D for count of each candidate. Compare | 3) Generate co-relations from the above statistics and scan for the count and compare the threshold count of hits |
|---|---|

3) Generate C2 candidates from L1. Scan D for count of each candidate. Compare

| Item | Support |
|------|---------|
| {1,2} | 3 |
| {1,3} | 2 |
| {1,4} | 3 |

candidate supp

3) Generate co-relations from the above statistics and scan for the count and compare the threshold count of hits

| Item | Hits |
|------|------|
| *ipadd->url* | 25 |
| *url->path* | 38 |
| *ipadd->path* | 27 |

---

4) Generate C3 candidates from L2 Scan D for count of each candidate and compare the candidate supp

| Item | Support |
|------|---------|
| {1,2,4} | 3 |
| {2,3,4} | 3 |

4) Generate further co-relations possible and compare the threshold count of hits

| Item | Hits |
|------|------|
| *ipadd->url->path* | 37 |

---

5) According to the pruning rule all the non empty itemsets should be the frequent item sets on this basis we find the final item set

5) According to the pruning rule all the non empty item sets should be the frequent item sets. On this basis we find the final item set There fore after pruning we get the final frequent item set as

*ipadd->url->path*

# 5. CONCLUSION AND FUTUTE SCOPE

In order to analyze educational log file, we have proposed the custom-built apriori algorithm. In particular, to find the different rules (co-relations), in a reasonable execution time, of all the frequent itemset from an educational log file. In order to perform the Web Usage Mining, the methodology that being introduced by Srivastava et al, [13] becomes major guide where it includes three main phases; data preprocessing, pattern analysis and pattern discovery. The simulated tests of our proposed algorithm, for which we have used parameters obtained from real runs of our educational web mining, are very promising. The rules (co-relations) resulted from our system helps the website developer in proper decision making and can improve their site effectively.

Many directions for future enhancements are open. Among them, we can mention:

1.  the adoption of more advanced techniques/tools such as fuzzy association rule mining algorithm using the interesting measures such as fuzzy support, fuzzy confidence, fuzzy interest and fuzzy conviction etc.
2.  the testing of the proposed work on a real distributed platform.

3. the use of the grid-computing paradigm to solve more challenging web mining problems and

4. this work primarily on the association rule mining but future research will be done using other data mining functions such as classification, clustering and so on.

## ACKNOWLEDGEMENTS

This work is partially supported by the Research Committee of the Guru Nanak Institutions (GNI), Hyderabad, India. The authors would like to thankfully acknowledge the support and cooperation of the teaching and non-teaching staff of Guru Nanak Institute of Technology, Hyderabad, India and College of Engineering, Osmania University, Hyderabad, India during this work.

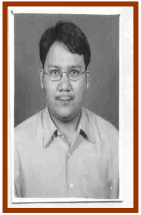

**Mr. Sandeep Singh Rawat** received his Bachelor of Engineering in Computer Science from National Institute of Technology - Surat (formerly REC - Surat), India and his Masters in Information Technology from Indian Institute of Technology, Roorkee, India. He is pursuing his Ph.D. at Osmania University, Hyderabad., India. He has presented three technical papers at international conferences and published paper in journals including IEEE Delhi Section and IEEE Computer Society Chapter, India.
His research interest includes Data Mining, High Performance Computing and Machine Learning.

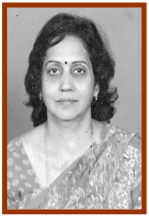

**Dr. Lakshmi Rajamani** is working as Professor and Head of the Department, Computer Science and Engineering, University College of Engineering, Osmania University, Hyderabad, India. She received M.Sc (Statistics) from IIT Kanpur, M.Phil (Computer methods) from University of Hyderabad and PhD (CSE) from Jadavpur University, Kolkata. She authored more than 25 papers in various National/International conferences and Journals. Her research interests are in the areas of Neural Networks, Artificial Intelligence, Distributed Computing & Data Mining.